\documentclass[runningheads]{llncs}
\usepackage{graphicx}
\usepackage{amsmath,amsfonts,bm}
\usepackage{graphicx}
\usepackage{caption}
\usepackage{subcaption}
\usepackage{graphicx}
\usepackage{float}
\usepackage{multirow}

\newcommand{\leftcenterhat}[1]{\expandafter\hat#1}

\usepackage[]{todonotes}

\begin{document}
\title{Identifying Patient Groups based on Frequent Patterns of Patient Samples}
%
%
\author{Seyed Amin Tabatabaei\inst{1}
\and
Xixi Lu\inst{2}
\and
Mark Hoogendoorn\inst{1}
\and
Hajo A Reijers\inst{2}
}
\authorrunning{S. Tabatabaei et al.}
%
\institute{Department of Computer Science, Vrije Universiteit Amsterdam \\
\email{\{s.tabatabaei, m.hoogendoorn\}@vu.nl}\\
 \and
Department of Information and Computing Sciences, Utrecht University\\
\email{\{x.lu, h.a.reijers\}@uu.nl}}
\maketitle              
\begin{abstract}
Grouping patients meaningfully can give insights about the different types of patients, their needs, and the priorities. Finding groups that are meaningful is however very challenging as background knowledge is often required to determine what a useful grouping is. In this paper we propose an approach that is able to find groups of patients based on a small sample of positive examples given by a domain expert.
Because of that, the approach relies on very limited efforts by the domain experts.
The approach groups based on the activities and diagnostic/billing codes within health pathways of patients. To define such a grouping based on the sample of patients efficiently, frequent patterns of activities are discovered and used to measure the similarity between the care pathways of other patients to the patients in the sample group. This approach results in an insightful definition of the group.
The proposed approach is evaluated using several datasets obtained from a large university medical center. The evaluation shows F1-scores of around 0.7 for grouping kidney injury and around 0.6 for diabetes.

\end{abstract}
\section{Introduction}
The application of machine learning techniques has become omnipresent in the health care domain over the last few decades. An important reason is the large amount of data that is collected in different parts and at different levels of health care systems. The huge amount of data is too complex for traditional methods to extract novel insights from. Machine learning techniques are able to transform this data into useful information for decision making by discovering patterns and trends \cite{koh2011data}.



One domain in which machine learning can play a role is in Patient classification systems (PCSs). PCSs provide a categorization of patients. Cases are usually categorized based on clinical data (i.e. diagnoses, procedures), demographic data (i.e. age, gender), and resource consumption data (i.e. costs, length of stay). Initially developed as an information management tool for clinicians to monitor quality and use of services, patient classification systems now, serve as a prospective payment system in many countries \cite{schreyogg2006methods}. While useful, such systems often do not align well with the patient groups as clinicians would define them. For example, patients getting reconstructive breast surgery caused by breast cancer or gender change could be merged into a single group, while their characteristics are completely different. Ideally, machine learning techniques would be able to find a suitable grouping that is clinically relevant and useful in practice. Furthermore, the definition of the group itself should be understandable for domain experts.

While the previously sketched scenario sounds ideal, it is far from trivial to establish. Machine learning can be used to cluster patients, but no guarantees can be given that this clustering is clinically relevant and acceptable for physicians. Secondly, there is a discrepancy between clinicians view on patients and the data that is stored about these patients (e.g. billing codes and interventions), making the task even more difficult~\cite{DBLP:journals/ijmi/BeekGK05}. Finally, any approach in this area should be sufficiently scalable.


In this paper, we present an approach to define patient groups which requires little input from domain experts.
The presented approach defines patient groups based on a small sample of example patients provided by medical experts. The approach uses frequent pattern mining to identify common characteristics among these example patients in the data (billing codes and activities related to the patient in our specific case), and uses them to create a definition of the patient group. 
An evaluation of the approach is performed on two real-life cases at a large academic hospital in the Netherlands, containing previously identified patient groups.

This paper is organized as follows. First, the precise problem we are trying to solve is explained in more detail in Section 2. Next, we describe our proposed approach (Section 3). We then present the experimental setup in Section 4, and the results in Section 5. Finally, Section 6 is a discussion.

\section{Problem Formulation}

In this section, we describe our research problem more in depth.
In the following, we first introduce preliminary concepts such as healthcare process and activities, events, traces, event logs, and patient groups. These form the basis for our research question and approach. Next, we use these concepts to define and discuss our research problem. 


\subsection{Preliminaries}
Each patient in a hospital follows a \emph{healthcare process}, which describes a set of \emph{healthcare activities} executed in a certain order to help treating the patient for a certain disease. 
An event log of patients is a set of \emph{traces}, each describing a patient pathway through the process. The executed activities result in such a \emph{trace} of events for the patient, where each \emph{event} records additional information regarding the executed activity. For example, Table \ref{table_example_database} shows a snippet of an event log of a healthcare process. Each row records an event executed, which contains information such as the event id, the patient id, the activity, the timestamps, the DBC code, and maybe some additional attributes regarding the event. 

Formally, let $A$ be the set of all possible activities and $D$ the set of all possible diagnostic codes. We have $N = \{n_1, n_2, \cdots, n_m\}$ patients. 
Let $L = \{ \sigma_1, \sigma_2, \cdots, \sigma_m \}$ be the event log of these patients, where $\sigma_i = \langle e_1, e_2, \cdots, e_{z_i}\rangle \in L$ is the trace of events of patient $n_i$; each event $e_j \in \sigma_i$ is a tuple $<a_j, d_j, t_j>$ where $a_j \in A$ refers to the healthcare activity of $e_j$, and $d_j \in D$ refers the diagnostic code of $e_j$, and $t_j$ refers to the time when $e_j$ is occurred. 

\subsection{Research Problem - Grouping Patients}
 In general, data recorded about the patients, the patients diagnosis, and the activities associated with the patient in the hospital are made available. We aim to cluster patients into meaningful groups by using (1) the original patient diagnostic classification (known in the Netherlands as Diagnostic and Billing Code (DBC)), (2) the activities recorded for each patients, and only (3) a small sample of patients provided by the medical expert. The sample contains examples of patients who should be grouped into one cluster. 

Many existing researches have conducted into various techniques to cluster the patients into a number of meaningful clusters (see e.g.,\cite{DBLP:conf/icdm/HiranoT13,DBLP:journals/midm/JayNGQ13,DBLP:conf/bpm/LakshmananRW13,DBLP:journals/is/RebugeF12}) by for example using healthcare or patient diagnostic codes. However, existing algorithms have difficulties handling the big data characteristics of the healthcare data (e.g., up to 6 thousands number of features and 120 thousands of patients to be clustered, see Section 4). 
Furthermore, they do not result in insightful definitions of the cluster that are in line with those of medical experts.

In this paper, we formulate the research problem as follows. Assuming a sample $P$ of the patients $N$ that should be clustered into one group $G$ is given, we would like to find all other patients in $N \backslash P$ that belong to this group $G$ with the highest recall and precision possible. In this way, we can also validate our results of each cluster independently with the experts and leverage the labeled patients. Additionally, we want the criteria for the grouping to be understandable for medical specialists.

Formally, given the healthcare activities $A$, the diagnostic codes $D$, the patient ids $N$, the event log $L$ of the patients $N$, and the set of patients $P \subset N$ that belongs to group $G$ labeled by the domain expert (the doctor), we would like to compute a set $\leftcenterhat{G} \subset N$ of patients, such that the set difference between $G$ and $\leftcenterhat{G}$ is minimized.

\begin{table}[!htb]

\centering
\caption{A snippet of an event log of a healthcare process}
\label{table_example_database}
\begin{tabular}{|c|c|c|c|}
 \hline
\, \textbf{Patient\_ID} \, & \, \textbf{Activity} \, & \, \textbf{Attribute} \, & \, \textbf{Date} \, \\ \hline
Patient1    & Action1  & DBC1 & Date1 \\ \hline
Patient2    & Action2  & DBC2 & Date1 \\ \hline
Patient3    & Action1  & DBC1 & Date3 \\ \hline
Patient1    & Action1  & DBC1 & Date2 \\ \hline
Patient2    & Action5  & DBC5 & Date2 \\ \hline
\end{tabular}
\end{table}

\section{Proposed Approach}

As explained before, only for a small sample of patients $P\subset N$ with $|P|<<|N|$ we know that they belong to the group $G$: $\forall_{p\in P}: p \in G$. For all other patients we do not know whether they belong to the group or not. Our challenge is to find a definition of the group based on the sample $P$ and the data we have available from this sample. Since we do not have any negative examples and due to the overwhelming number of features, we have decided to apply frequent pattern analysis as a first step, followed by a further refinement of the definition in a second step. Finally, we determine how to apply the found definition to identify the group members.

\subsection{Finding Frequent Patterns}

Let us start with the definition of a frequent pattern. A frequent pattern is a pattern (a set of items, subsequences, sub graphs, etc.) that occurs in a dataset with frequency no less than a user-specified threshold. The approach originated from market basket analysis and was first proposed in \cite{agrawal1993mining}. In our case, we intend to find frequent patterns among the activities conducted around a patient (i.e. from the set $A$) as they are the key identifiers of the patient groups according to the medical experts. Assume that $A_p$ is the set of activities that occurs in the event log of patient $p$: $\forall a_i \in A_p: \exists e_j \in \sigma_p: e_j = <a_i, -,->$. As said, we are looking for frequent patterns across the different patients. Such frequent patterns can contain one or more activities $F=\{a_1, \dots, a_k\}$. Frequent patterns have a certain support, representing among what fraction of the patients in sample $P$ the pattern occurs:
\begin{align}
support(F) = \frac{|\{p|F \subset A_p\}|}{|P|}
\end{align}
\noindent Patterns are only considered when they meet a certain support threshold $\phi_a$, i.e. $support(F) \geq \phi_a$.

The most well-known algorithms for finding frequent patterns are Apriori~\cite{agrawal1994fast} and FP-growth~\cite{han2000mining}. We have selected FP-growth since its run time increases linearly with the number of patients and activities, while for Apriori the run time increases exponentially depending on the number of activities. Running the algorithm results in a set of frequent patterns. We then select the most specific pattern (i.e. the pattern with the highest value for $k$) from this set. We select only one pattern to make the definition of the group understandable for domain experts and the longest frequent pattern is the most specific characterization of the group. We will refer to it as $F_a$ from now onwards. We have now obtained a way to define the group, however merely activities showed to be insufficient to identify the group completely. We therefore add additional criteria.

\subsection{Finding Additional Criteria}

To make the patterns more specific, we add criteria related to the DBC code (remember the example given in the introduction for the breast cancer and gender patient, these could at this point still be categorized under the same group). We add the DBC code in a bit different way, namely by how often the activities in the selected frequent pattern and the billing codes coincide. Assume $D_p$ is the set of DBC codes in the logs of patient $p$: $\forall d_i \in D_p: \exists e_j \in \sigma_p: e_j = <-,d_i,->$.
For each $d$: $d \in D_p , p \in P$, 
we compute the fraction of patients from sample $P$ for which $d$ coincides with at least one element of our initial frequent pattern $F$ as these are the core mechanism to select the patients on:
\begin{align}
support(d) = \frac{|\{p|b \in B_p \ , \exists e_j \in \sigma_p , \exists a_k \in F_{a}: e_j= <a_k,b,->\}|}{|P|}
\end{align}

\noindent If the support exceeds a certain threshold $\phi_d$, we add the code to the set of DBC codes $D$. Hence, $D = \{d| d \in D_p, support(d) \geq \phi_d\}$.

After these two steps we end up with a frequent pattern $F$ for activities and a set of selected DBC codes $D$.

\subsection{Classification}

Now we can score all the patients $p$ based on how far they satisfy our two newly defined criteria. Here, 0 is the optimal score, showing no discrepancy between the patient and the definition of the group.
\begin{align}
activity\_score(p) = |F| - |\{a_i|a_i \in A_p, a_i \in F\}|\\
dbc\_score(p) = |D| - |\{d_i|d_i \in D_p, d_i \in D\}|
\end{align}

Combining these into one measurement is not a trivial task, neither is determining when a patient is still in the group.
In the next part, we will explain how we determine the cut-off based on $P$.

\subsection{Determining cut-off values}

The grouping of our patients depends on both the support thresholds ($\phi_a$ and $\phi_d$) used to find patterns, and cut-off values for how many of the found criteria should meet to be considered part of the group.

For the support thresholds, we assume a domain expert sets the support threshold. To do that, we start with $\phi_a=1$ and decrease its value with a small step size. By decreasing the value of this parameter, the constraints on the frequent patterns become more relaxed and new activities are added to $F_a$. In each step, the expert can study the new activities added to $F_a$, and see if they can be good features of the input sample group. The procedure will stop when the expert finds new activities not being representative for the sample group. Then, $\phi_d$  should be adjusted by the same strategy. As said before, the frequent DBCs are dependent on $F_a$ (the opposite dependency is not the case). That is why $\phi_d$ should be adjusted after $\phi_a$ .

For the cut-off values, let us define the precise meaning of these first. A patient is defined as part of the group $\hat{G}$ when it remains under or is equal to the cut-off value for both criteria we have explained before:
\begin{align}
\leftcenterhat{G_{\alpha_F, \alpha_D}} = \{p|p \in N, activity\_score(p) \leq \alpha_F, billing\_score(p) \leq \alpha_D\}  
\end{align}
\noindent The cut-off values are referred to as $\alpha_F$ and $\alpha_D$. When we are optimizing the grouping (and thus these parameters) based on our small sample we would like to optimize both the precision and recall, and hence, use the F-measure:
\begin{align}
precision_{\alpha_F, \alpha_D} = \frac{|\leftcenterhat{G_{\alpha_F, \alpha_D}} \cap G|}{|\leftcenterhat{G_{\alpha_F, \alpha_D}|}}
\end{align}
\begin{align}
recall_{\alpha_F, \alpha_D} = \frac{|\leftcenterhat{G_{\alpha_F, \alpha_D}} \cap G|}
{|G|}
\end{align}
\begin{align}
Fn\_measure_{\alpha_F, \alpha_D} = (1+n^2) \cdot  \frac{precision_{\alpha_F, \alpha_D} \cdot recall_{\alpha_F, \alpha_D}}
{(n^2 \cdot precision_{\alpha_F, \alpha_D}) + recall_{\alpha_F, \alpha_D}}
\end{align}
Do not having access to negative examples, making it near impossible to optimize our parameters on this. We therefore focus on the recall in this stage, and focus on getting the highest recall among the group of patients we know should belong to the group (i.e. from our sample $P$):
\begin{align}
\overline{recall}_{\alpha_F, \alpha_D} = \frac{|\leftcenterhat{G_{\alpha_F, \alpha_D}} \cap P|}{|P|}
\end{align}
Assuming that $|P|$ is a representative sample, the expected value of $\overline{recall}$ is equal to the real value of the recall. Of course, just optimizing on the $\overline{recall}$ is likely to give us a very loose definition, and hence a low precision. We therefore use an elbow based method to identify the point where the recall increase starts to flatten, and select that as the cut-off point.
An alternative criteria which is suggested in \cite{lee2003learning}, is to find the set of parameters which maximize $\frac{\overline{recall}^2 }{ |\hat{G}|}$.

\section{Experimental Setup}

\begin{figure}[]
\centering
\begin{subfigure}{0.33\textwidth}
\centering
\includegraphics[width=0.9\textwidth]{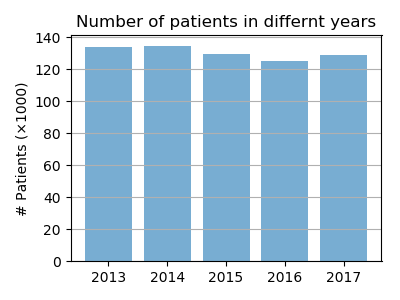}
\caption{\#Patients each year  }
\label{fig_Dataset_a}
\end{subfigure}
\begin{subfigure}{0.33\textwidth}
\centering
\includegraphics[width=0.9\textwidth]{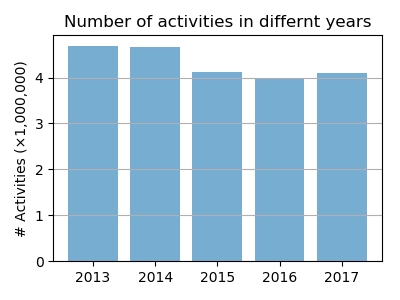}
\centering
\caption{\#Activities each year}
\label{fig_Dataset_b}
\end{subfigure}%
\begin{subfigure}{0.33\textwidth}
\centering
\includegraphics[width=0.9\textwidth]{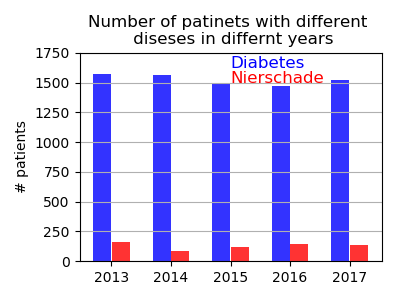}
\centering
\caption{Size of ground truths}
\label{fig_Dataset_c}
\end{subfigure}%

\caption{General information over dataset and ground truth}
\label{fig_Dataset}
\end{figure}

We evaluate the proposed approach on anonymized records of patients of the VU University Medical Center Amsterdam, collected between 2013 and 2017. This dataset contains the event logs of the patients in line with the description in Section 2. In total 329,783 patients are present in the dataset. There are 8,360 unique activities and 2,273 unique DBCs. In total more than 35 million activity events are recorded in the dataset. Figure~\ref{fig_Dataset_a} shows the number of patients and Figure~\ref{fig_Dataset_b} shows the number of activities in different years.

In addition to the event logs, lists of patients for two groups that were defined by medical experts (diabetes and kidney injury) in each year were provided. The number of patients in these groups ($G_{diabetes}$ and $G_{kidney}$) over different years is shown in Figure~\ref{fig_Dataset_c}. These groupings are assumed to be the ground truth. According to the domain experts, labelling approximately 30 patients would be a feasible task. Hence, we take a random sample of 30 patients for both groups ($P_{diabetes}$ and $P_{kidney}$, with  $|P_{diabetes}|=|P_{kidney}|=30$. To study the validity of the parameter settings and the impact on the quality of the grouping, we split this sample in a 50/50 way again, using 15 patients to generate our definitions of the set. The remaining patients in the group, but outside of this sample of 30 is left for calculating $\overline{recall}$. Finally, we set the support threshold to $\phi_a=\phi_d= 0.8$ based on preliminary experiments with domain experts. Furthermore, we use the F1-score as our final evaluation metric.

\section{Results}

Let us move to the results. First, we will present the results related to the frequent patterns, followed by the optimization of the parameters. We end with the F1-scores we obtain.

\subsection{Frequent Patterns}

The first step in this approach is finding frequent patterns of activities and the relevant DBCs for the training samples of the two groups under investigation. Table \ref{table_FPs_Diabetes_2017} shows the sets $F_{diabetes}$ and $D_{diabetes}$ that were found based on a sample from 2017. For the sake of brevity, we do not show the table related to kidney injury.
\begin{table}[h]
\caption{Frequent activities and DBC codes for Diabetes in 2017.}
\begin{tabular}{|p{8.5cm}|p{3.2cm}|}
\hline
\textbf{$F_{diabetes}$} & \textbf{$D_{diabetes}$} \\\hline

{\scriptsize first polyclinic visit} &  {\scriptsize diabetes mellitus without} \\
{\scriptsize hemoglobin (including extra measurements)} &{\scriptsize secondary complications} \\  
{\scriptsize hemoglobin A1 blood}& \\                              
{\scriptsize potassium. }& \\                                                
{\scriptsize creatinine}& \\                                            
{\scriptsize blood collection and chemical and microbiotic analysis }& \\ 
{\scriptsize phone consult}& \\  
\hline
\end{tabular}
\label{table_FPs_Diabetes_2017}
\end{table}
\subsection{Classification of Sample Patients}
\begin{figure}[!htb]
\centering
\begin{subfigure}{0.45\textwidth}
\centering
\includegraphics[width=5.75cm]{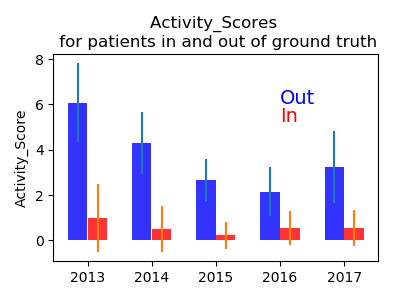}
\caption{Diabetes}
\label{fig_missed_FPs_a}
\end{subfigure}
\begin{subfigure}{0.45\textwidth}
\centering
\includegraphics[width=5.75cm]{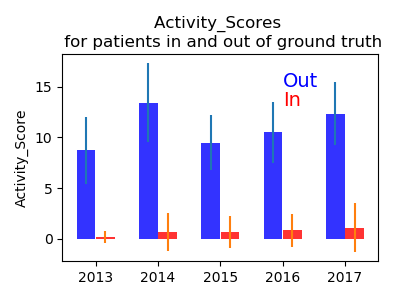}
\centering
\caption{Kidney injury}
\label{fig_missed_FPs_b}
\end{subfigure}%
\caption{Average number of missed frequent activities for patients in and outside of $G$}
\label{fig_missed_FPs}
\end{figure}

For each patient in the dataset, two distances from the training sample group are calculated as explained in the approach (i.e. $activity\_score$ and $dbc\_score$).
Figure~\ref{fig_missed_FPs} shows the average  $activity\_score$ for the patients inside and outside the sample of labeled patients used to generate the patterns for both use cases. As depicted in Figure~\ref{fig_missed_FPs}, there is a significant difference between the distances of patients inside and outside ground truth. This shows that patterns have been found that distinguish the group of patients from other patients.

\subsection{Classification of All Patients}
\begin{figure}[!htb]
\centering
\begin{subfigure}{0.32\textwidth}
\centering
\includegraphics[width=3.5cm]{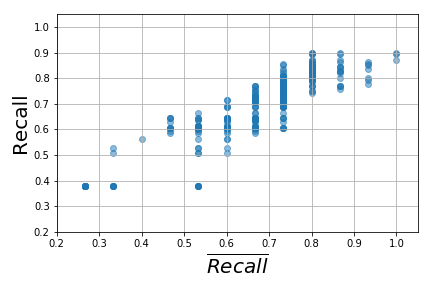}
\caption{$\overline{recall}$ VS. recall}
\label{fig-Fmeasure_a}
\end{subfigure}
\begin{subfigure}{0.32\textwidth}
\centering
\includegraphics[width=3.5cm]{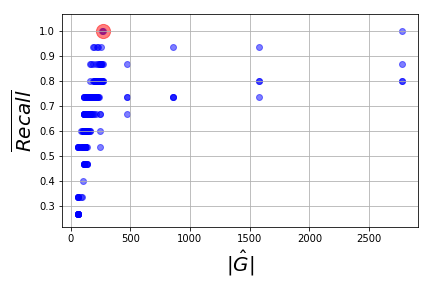}
\caption{Number of selected patients VS. $\overline{recall}$}
\label{fig-Fmeasure_b}
\end{subfigure}
\begin{subfigure}{0.32\textwidth}
\centering
\includegraphics[width=3.5cm]{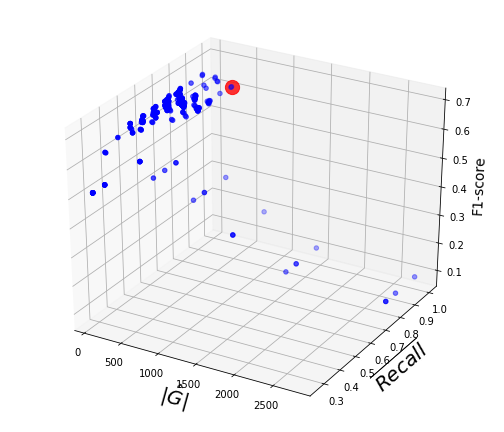}
\centering
\caption{$F1-score$}
\label{fig-Fmeasure_c}
\end{subfigure}%

\caption{
Graphs related to kidney injury in 2017:
(a) relationship between $\overline{recall}$ and real recall, (b) recall versus the number of patients included in $\hat{G}$, and (c) $\overline{recall}$ and number of selected patients versus the F1-score.
}
\label{fig-Fmeasure}
\end{figure}

Now that we have found patterns that seem to distinguish the patients, we are still faced with finding proper cut-off values. For this we use the estimation of the recall based on our small sample. Figure~\ref{fig-Fmeasure_a} shows how the value of $overline{recall}$ computed over the 15 patients in the test set compares to the true recall (which we can compute for this study as we have access to the full grouping). Spearman's correlation test shows a strong correlation between these the recall on the small sample and the overall recall (correlation=0.87, pvalue=6.9e-47).
%
%
%
As explained before, we select the optimal values for the parameters $\alpha_F$ and $\alpha_D$ based on the elbow of the recall. Figure~\ref{fig-Fmeasure_b} shows that in general by increasing the size of selected group, $|\hat{G}|$, the value of $\overline{recall}$ is increasing. However, a clear turning point can be seen where the $\overline{recall}$ does not increase significantly with a substantial increase of $|\hat{G}|$. The accompanying parameter values are selected. In Figure~\ref{fig-Fmeasure_b}, and Figure~\ref{fig-Fmeasure_c}, this point is highlighted in red. Note again that we do not use any additional information besides the sample to determine this.

Following this approach for the two diseases and years, we evaluate how suitable the grouping is given the ground truth of all patients. As can be seen, the maximum achieved F-measure for kidney injury is always higher than the value for diabetes. One reason is that although the number of patients with diabetes is higher than number of patients with kidney injury (Figure~\ref{fig_Dataset_c}), the size of the training sample group is the same. Hence, it could potentially be a too diverse group. For example in year 2017, more than 10\% of the whole patients with kidney injury is used as training set, while this percentage is less than 1\% for Diabetes. Still, the results on the F1-score are within reasonable bounds according to domain experts.
\begin{figure}[!htb]
\centering
\includegraphics[width=0.7\textwidth]{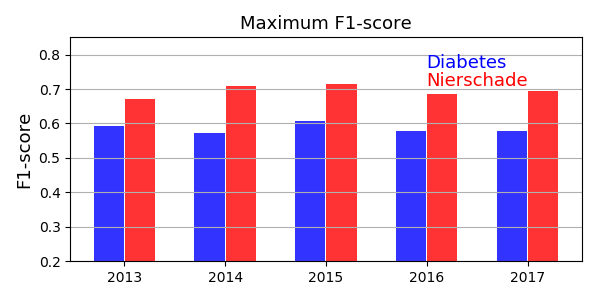}
\caption{Maximum achieved F-measure for two diseases in different years. In each case, a sample group of 15 patients is used as training set.} 
\label{fig-Maximum_fmeasure}
\end{figure}
\section{Conclusion and Future Work}

We have presented an approach that is able to extract a definition of a group of patients based on a small sample of example patients given by a medical expert. Hereto, frequent item sets have been exploited, and extended to make them suitable for the case at hand. The approach results in an insightful definition of the group. An experimental evaluation of the approach based on a real life dataset with two groups defined by medical experts shows that the approach is able to generate reasonable F1-scores. The proposed approach is very feasible since it puts a minimum load on the domain experts. Moreover,
The approach is flexible in that it allows for varying strictness of the group definition by varying the parameters of the approach.

For future work, we plan on extending the selection mechanism to make it more robust and improve the F1-scores even further. Furthermore, we want to apply the approach to more rare cases and involve medical experts more heavily in the process.

\bibliographystyle{splncs04}
\bibliography{AIME2019bibliography}
\end{document}